# Single Photoelectron Detection after Selective Excitation of Electron-Heavy Hole and Electron-Light Hole Pairs in Double Quantum Dots

K. Morimoto,[1] T. Fujita,[1] G. Allison,[1] S. Teraoka,[1] M. Larsson,[1] H. Kiyama,[1] S. Haffouz,[2] D. G. Austing,[2] A. Ludwig,[3] A. D. Wieck,[3] A. Oiwa,[1] and S. Tarucha[1,4]

[1] *Department of Applied Physics, The University of Tokyo, 7-3-1 Hongo, Bunkyo-ku, Tokyo 113-8656, Japan*
[2] *National Research Council of Canada, M50, Montreal Road, Ottawa, Ontario, Canada K1A 0R6*
[3] *Lehrstuhl für Angewandte Festkörperphysik, Ruhr-Universität Bochum, Universitätsstraße 150, Gebäude NB, D-44780 Bochum, Germany*
[4] *Center for Emergent Matter Science (CEMS), RIKEN, 2-1 Hirosawa, Wako-shi, Saitama 351-0198, Japan*
(Dated: January 17, 2014)

We demonstrate the real-time detection of single photogenerated electrons in two different lateral double quantum dots made in AlGaAs/GaAs/AlGaAs quantum wells having a thin or a thick AlGaAs barrier layer. The observed incident laser power and photon energy dependences of the photoelectron detection efficiency both indicate that the trapped photoelectrons are, for the thin barrier sample, predominantly photogenerated in the buffer layer followed by tunneling into one of the two dots, whereas for the thick barrier sample they are directly photogenerated in the well. For the latter, single photoelectron detection after selective excitation of the heavy and light hole state in the dot is well resolved. This ensures the applicability of our quantum well-based quantum dot systems for the coherent transfer from single photon polarization to single electron spin states.

The rapid escalation of information processing and storage in the last decade has necessitated the development of novel technologies to ensure highly secured long distance communication. With recent advances in theoretical and experimental studies, quantum information is believed to be a potential candidate for future technology with absolutely secure information exchange. Major challenges in quantum information systems include long distance communication and multi-node networks, which require coherent coupling of photons to solid-state quantum bits (qubits) as used for implementing quantum memory and quantum entanglement. They are both fundamental elements for constructing a quantum repeater, distributed quantum computing and hybrid quantum information network [1].

A noticeable proposal for the coherent transfer from photon-polarization qubits to electron spin qubits based on the optical selection rule [2] has been experimentally demonstrated for an ensemble of photons and electrons in g-factor engineered GaAs quantum well (QW) systems [3,4]. However, practical use of a quantum information network requires quantum state transfer from a *single* photon to a *single* electron spin, which is feasible in a quantum dot (QD) with three-dimensionally confined electrons [5-7]. Electrically gated QDs have an advantage that the confined electron spin can be manipulated and detected for making various kinds of qubit gates. In this context electron spin is an appropriate partner for photons. For the first step toward the verification of coherent transfer between single quanta, A. Pioda *et al.* [8] realized the real-time detection of trapping and resetting of single photogenerated electrons using a nearby quantum point contact (QPC) as a charge sensor. The change of electron number in the single QD was clearly monitored by measuring the change of the QPC conductance when the photogenerated electron escaped from the dot. The authors proved that the photoelectron trapping can be electrically detected within a time significantly shorter than the spin-flip time $T_1$. This technique was later extended to double QDs (DQDs) and the improved ability to detect single photoelectrons as well as their spin orientations through inter-dot tunneling has been demonstrated [9]. Note in DQDs spin-related phenomena such as Pauli spin blockade, and electron spin resonance can be utilized for nondestructive spin readout, and coherent spin rotation, respectively [10-13].

Despite the previous innovative results, those samples are not suitable for coherent transfer. The proposed coherent transfer scheme postulates specified systems like QWs with a GaAs well between two AlGaAs barriers. The heavy and light hole bands are energetically separated in the QW and the electron and light hole g-factors can be tuned to satisfy the condition of coherent photon to spin information transfer: $g_e\mu_B B \ll \Delta E_{ph} \ll g_{lh}\mu_B B$ [2,4]. Here, $g_e$, and $g_{lh}$ are the g-factors of conduction band electron, and valence band light hole, respectively, $\mu_B$ the Bohr magneton, and $\Delta E_{ph}$ the photon energy bandwidth. Selective excitation of electron-heavy hole and electron-light hole in the QW is the key ingredient, since the coherent transfer is only possible with excitation of nondegenerate heavy or light hole band under magnetic fields orthogonal to the light propagation direction, producing electron spin superposition state which is disentangled from the hole spin state [2]. Single photoelectron trapping in a QD formed in a QW with a weak confinement was previously studied but the electron-heavy hole and electron-light hole excitation could not be well resolved [14]. We could detect the resonant excitations of single photoelectrons from distinct quantum states in a QD formed in a relatively narrow QW. In this work we investigate in detail the single photon response for such selective excitation in the DQDs. By comparing the photoelectron detection efficiencies between two DQDs made in a QW with a thin or a thick lower barrier layer and measuring the photon energy dependence of the detection efficiency we demonstrate the trapping of single photoelectrons resonantly excited from the heavy and light hole states as an important prerequisite for the photon to spin coherent transfer.

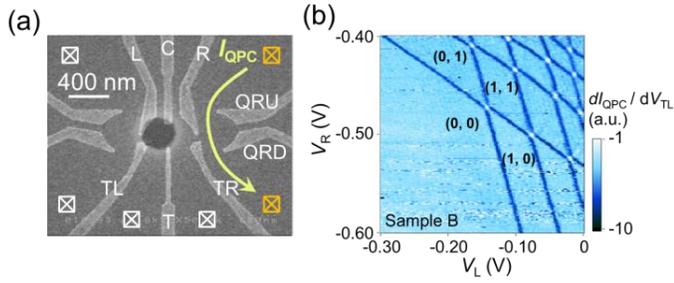

FIG. 1. (Color online) (a) A composite scanning electron micrograph of sample B. The charge states of the DQD are monitored by the current $I_{QPC}$ through the nearby QPC. A thick Ti/Au mask is deposited on top of the wafer with a 400 nm aperture placed above the left dot for selective photon irradiation. The metal mask and surface gates are insulated from each other by a 100 nm calixarene dielectric layer. (b) Typical charge stability diagram measured for sample B with an AC charge sensing technique.

Lateral QDs were fabricated on two kinds of AlGaAs/GaAs/AlGaAs QW wafers. Both wafers have the same layer sequence except for the AlGaAs barrier layers: a 5 nm GaAs capping layer, a 65 nm Si doped AlGaAs layer, a 30 nm undoped AlGaAs spacer, a 7 nm GaAs well, an undoped AlGaAs barrier, and a thick GaAs buffer layer. The upper (lower) barrier is 95 nm $Al_{.265}Ga_{.735}As$ (20 nm $Al_{.265}Ga_{.735}As$) for wafer I, while it is 95 nm $Al_{.34}Ga_{.66}As$ (100 nm $Al_{.34}Ga_{.66}As$) for wafer II. A two-dimensional electron gas is only formed in the GaAs QW and no parallel conduction was observed in both Hall and Shubnikov-de Haas measurement.

Gate-defined DQDs were fabricated using wafer I, named 'sample A' and wafer II, named 'sample B', respectively (see Figure 1(a)). The following experiments were carried out in a 0.4 K $^3$He cryostat with an optical window, through which light emitted from either a 780 nm pulsed semiconductor laser diode or a wavelength tunable Ti:Sapphire laser was irradiated onto the sample. The focused laser beam diameter at the sample surface was tuned from 150 to 300 μm for each experiment, and the emitted laser pulse was selectively irradiated to the left dot through a 400 nm diameter aperture made in a 300 nm-thick Au mask. Figure 1(b) is the charge stability diagram of sample B, measured before the photoelectron-trap experiment. The typical honeycomb structure represents the formation of a DQD. The (0, 0) charge state is clearly identified.

In the photoelectron-trap experiment we must consider two types of processes depending on the thickness of the lower barrier. This arises from the difference in the photoexcitation as shown schematically in Figs. 2(a) and (b). Assuming that the incident photon energy is larger than the GaAs bandgap and smaller than the AlGaAs bandgap, the irradiated photons is absorbed in either the well or buffer. Analogous to the HEMT case of ref. 8, a photoexcited electron-hole pair in the buffer region is spatially separated by the internal built-in electric field, and the photoelectron can tunnel through the lower barrier to reach the well if the barrier is thin enough (see Fig. 2(a)). In contrast, the photoelectron is filtered out by a thick barrier. As a result, the dominant photoelectron trapping process is expected to arise solely from the direct excitation of an electron-hole pair in the QW (see Fig. 2(b)).

With regard to the photon to spin quantum conversion, single photoelectron trapping after selective excitation of electron-heavy hole and electron-light hole pairs in the QW has to be resolved [2]. The comparison of photoelectron trapping events in the two QW samples with different lower barrier thicknesses may give us perspectives of how to resolve the selective photoexcitation. First, the photon detection efficiency is smaller in the thin lower barrier sample than in the thick lower barrier sample, since both types of excitation occur in sample A, while direct excitation of the QW predominantly contributes in sample B. Second, one should observe strong photon energy selectivity in the photon response for sample B due to the strongly enhanced absorption efficiency at particular photon energies of the resonant electron-heavy hole (or light hole) excitation. These characteristics are experimentally accessible by extracting the statistical probability of single photoelectron trapping averaged over many single-shot measurements as a function of laser power or photon energy.

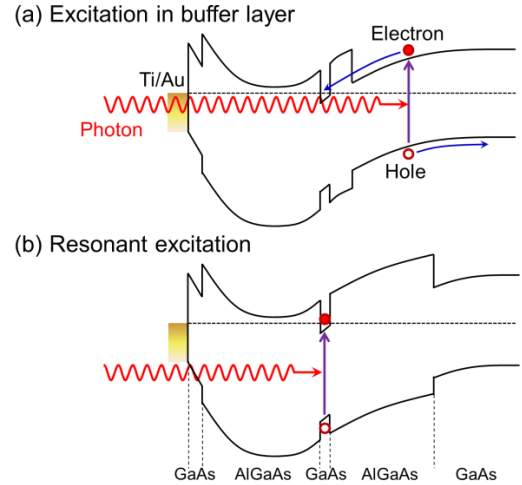

FIG. 2. (Color online) Schematic illustrations of two mechanisms of photoelectron excitation in AlGaAs/GaAs/AlGaAs QW wafers. (a) Excitation of an electron-hole pair in the GaAs buffer layer and subsequent transfer of photoelectron to the QW. (b) Resonant excitation of an electron-hole pair in the QW.

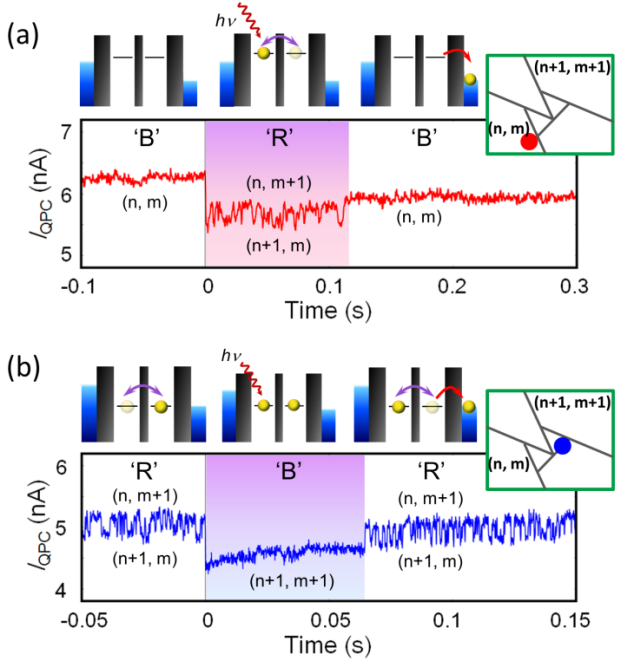

FIG. 3. (Color online) Photon response in the charge sensor current measured for sample A at two different bias points as schematically shown by the red circle (a) and the blue circle (b) in the stability diagram in the inset. Picosecond semiconductor laser pulses are used for photon irradiation onto the dot. The colored region in each figure shows the time duration of photoelectron trapping in the DQD. 'B', and 'R' stands for 'blocked', and 'resonant' inter-dot tunneling in the weakly coupled DQDs, respectively. The dot-lead, and inter-dot tunnel rates are tuned to about 10 Hz, and 500 Hz, respectively. The source-drain bias $V_{sd}$ is set to 1 mV. An inter-dot tunneling of a single photoelectron induces the fluctuating signal in the sensor current in (a). In (b) inter-dot tunneling of an initially trapped electron is impeded by Coulomb repulsion from the added photoelectron. The energy level profiles of the DQD and electron dynamics are depicted in the upper side of each figure.

To investigate the two types of photoelectron trapping processes, we performed photon irradiation measurements with few electrons in the DQDs. In this setup, photon-charge conversion events are observed as abrupt jumps in the charge sensor current, $I_{QPC}$, corresponding to the change of the electron number in the DQD. The added photoelectron escapes from the dots on a time-scale of the dot-lead tunneling, resulting in the recovery of the shifted current.

Figures 3(a) and (b) show typical single photon responses for sample A measured at two different gate bias points in the stability diagram. In Fig. 3(a) the DQD is initialized in the (n, m) state with two energetically aligned empty states. When a single photoelectron is trapped in the DQD upon photon irradiation at $t = 0$, the QPC current suddenly starts fluctuating between two values. The fluctuation rate is equal to the inter-dot tunneling rate, which is evaluated from separate real-time detection of single electron tunneling experiments, indicating repeated transition between the (n+1, m) and (n, m+1) states. The current fluctuation finally stops when the photoelectron escapes from the DQD to the lead. The QPC current after the photoelectron escape is smaller than the level before the photon irradiation, probably due to persistent photoconductivity induced by photon absorption outside of the metal mask.

On the other hand, in Fig. 3(b) the single photon response is opposite to that in (a). We observe QPC current fluctuations before the photon irradiation and after the photoelectron escape, but no current fluctuation in between. This indicates that the (n+1, m) - (n, m+1) inter-dot tunneling is stopped by filling the (n+1, m+1) state. The gradual background current change clearly seen in this region reflects temporal charge redistribution near the QPC sensor [8].

Therefore two different methods can evenly apply for single photon detection in the DQD: 'initially-blocked' (IB) and 'initially-resonant' (IR). Only the 'IB' method was studied in our previous work [9] but here we find that the 'IR' method enables robust and nondestructive single photon detection as well.

We further measured the counting probability of the signals for the two detection methods as a function of incident laser power. The result is shown in Fig. 4. Red circles (blue squares) are the probabilities of detecting single photoelectron trapping by the 'IB' ('IR') method. The incident photon flux along the horizontal axis is determined by estimating the number of photons incident on the dot through the aperture mask (see the supplementary information in ref. 9). For both methods, we observe a monotonic increase of the counting probability with increasing incident photon flux as critical evidence that the observed photon response indeed arises from photogenerated electrons in the dot. In addition, almost identical power dependence observed for the methods IR and IB proves that both of them are equally feasible for single photon detection in DQDs. The linear fitting by the black solid line gives a photon detection efficiency of 20±5% in this system. This value is comparable to the value of 15, and 17% in ref. 8, and 9, respectively.

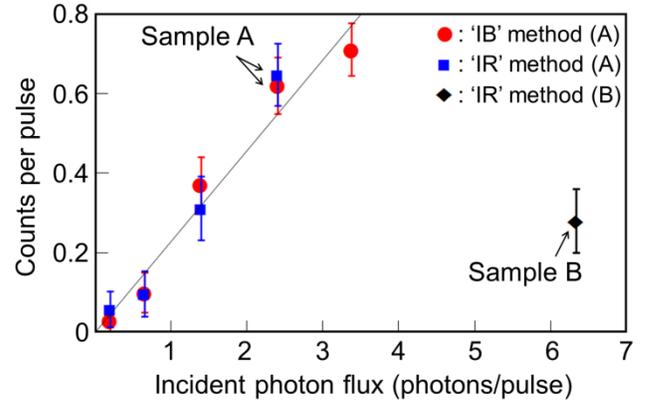

FIG. 4. (Color online) Laser power dependence of the single photon counting probability in sample A and B. Red circles and blue squares show the counting probabilities in sample A measured using different methods depicted in Figs. 3(a) and (b). The photon energy is fixed at 1.59 eV. The solid line is the linear fitting for sample A. The black diamond is the photon counting probability in sample B measured with the photon energy of 1.578 eV.

The photon absorption rate in a 7 nm GaAs QW is calulated to be about 1% using the photon absorption coefficient in bulk GaAs ($\sim 10^4$ cm$^{-1}$). The unexpectedly high efficiency obtained for sample A can be ascribed to the counting of photoelectrons generated in the GaAs buffer layer as discussed in Fig. 2(a), as well as those created in the well. The time scale of an electron tunneling through the 20 nm lower barrier is calculated to be the order of a microsecond using WKB approximation, where a rectangular potential is assumed. Our sensor time resolution (100 μs) is much larger than the tunneling time. Therefore most of the electrons are photogenerated in the GaAs buffer and then tunnel into the QW or dot and are unresolvable from those directly photogenerated in the dot. The calculation also indicates that for the 100 nm Al$_{34}$Ga$_{66}$As barrier the tunneling time is much larger than the measurement time scale, so that photogenerated electrons in the dot only contribute to the measured photon counting.

The black diamond in Fig. 4 shows the measured photon counting probability in sample B using a Ti:Sapphire laser (1.578 eV). Assuming a linear dependence of the counting number on the incident photon flux, the efficiency is evaluated to be 4.5%. This value is considerably smaller than that for sample A, and therefore indicates that the 100 nm thick lower barrier is thick enough to suppress the electron tunneling from the buffer region to the dot. The larger efficiency than the above estimate of 1% might be due to the 2D excitonic effect in the QW. Multiple reflections inside the sample due to the thick metal mask on top of the QD may also enhance the efficiency.

We performed the same experiment for sample B using a Ti:Sapphire laser with a fixed incident photon flux. Figure 5 shows the single photoelectron counting probability as a function of photon energy. A single laser pulse includes an average of 6.3 incident photons on the dot through the aperture, and the counting probability is determined from 100 to 200 photon pulse irradiations. A significant number of photoelectron counts are only observed for photon energies larger than 1.57 eV, which is well above the gap of bulk GaAs and consistent with the renormalized gap calculated for the QW.

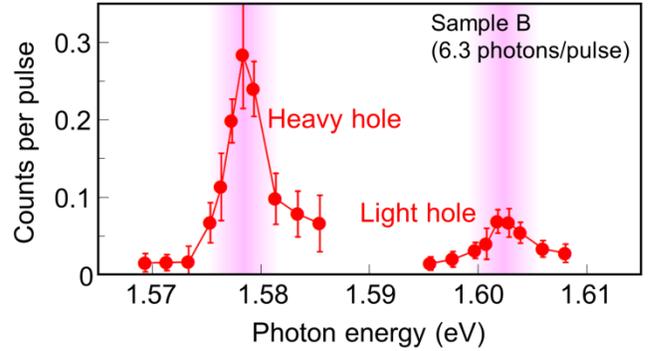

FIG. 5. (Color online) Photon energy dependence of single photon counting probability in sample B. Each data point in red is statistically deduced after out of 100 to 300 laser pulse irradiations. Peaks at 1.578 eV and 1.602 eV indicate resonant excitation of the heavy hole state and the light hole state, respectively.

Therefore, the observed photon absorption is considered to occur predominantly in the dot. We observe an unambiguous peak of the counting probability at 1.578 eV. Consistently in photoluminescence experiments of a piece of wafer II we observe a peak at a similar photon energy (not shown). Therefore we assign the peak at 1.578 eV in Fig. 5 to heavy hole excitation in the QW. This assignment is also supported by the fact that no other peaks are found at smaller photon energies. Moreover, there is a small but definite peak at 1.602 eV which is likely due to light hole excitation. The separation between the two peaks is consistent with our 1D-calculation of 20 to 30 meV for the heavy and light hole separation [15]. The photon absorption ratio between the heavy hole excitation and the light hole excitation is expected to be 3:1 [16] in GaAs. This is qualitatively consistent with the ratio of the observed peak heights. Note that the peak broadening in Fig. 5 might be caused by the inhomogeneity of the well thickness or spatially distributed peak position by the non-uniform electric field from the surface gates. This broadening makes it difficult to resolve the effect of lateral confinement in the QD since the level separation due to the lateral confinement is smaller than 1 meV [5].

In conclusion, we demonstrated single photon detection in the DQDs made in an AlGaAs/GaAs/AlGaAs QW with a thin or thick barrier layer. The laser power dependence of the photon response in these samples revealed that in the thinner barrier sample photoelectrons are predominantly generated in the buffer layer, whereas they are only generated in the QW for the thicker barrier sample. A striking photon energy dependence of the photoelectron counting probability due to selective excitation of the heavy and light hole state is observed in the thicker barrier sample. The selective excitation observed here ensures feasibility of QW-based lateral dot devices applied to quantum media conversion from single photon to single electron spin states.

This work was supported by Grants-in-Aid for Scientific Research A (No. 25246005), Innovative area (Grant No. 21102003), FIRST program, QuEST Grant No. (HR-001-09-1-0007), IARPA, MEXT Project for Developing Innovation Systems, and QPEC, The University of Tokyo. T. F. is supported by JSPS Research Fellowships for Young Scientists. A. L. and A. D.W. acknowledge gratefully support from DFG-SPP1285, BMBF QuaHL-Rep 16BQ1035, and DFH/UFA CDFA-05-06.

---